\documentclass[twocolumn,showpacs,preprintnumbers,amsmath,amssymb,superscriptaddress,UTF8]{revtex4}
\usepackage{dsfont}
\usepackage{cases}
\usepackage{graphicx}
\usepackage{dcolumn}
\usepackage{bm}
\usepackage{amsmath}
\usepackage{hyperref}
\usepackage{cleveref}

\UseRawInputEncoding

\begin{document}
\title{Modulating quantum evolution of moving-qubit by using classical driving}

 \author{Qilin Wang}
 \thanks {These authors contribute equally to this article.}
 \affiliation{Synergetic Innovation Center for Quantum Effects and Application, Key Laboratory of Low-dimensional Quantum Structures and Quantum Control of Ministry of Education, School of Physics and Electronics, Hunan Normal University, Changsha, 410081, People's Republic of China.}
\author{Jianhe Yang }
\thanks {These authors contribute equally to this article.}
\affiliation{School of Physics and Electronics, Central South University, Changsha 410083, People's Republic of China.}
\author{Rongfang Liu}
\thanks{These authors contribute equally to this article.}
\affiliation{School of Physical Science and Technology, Lanzhou University, Lanzhou 730000, China.}
\author{Hong-Mei Zou \thanks}
\email{zhmzc1997@hunnu.edu.cn}
\affiliation{Synergetic Innovation Center for Quantum Effects and Application, Key Laboratory of Low-dimensional Quantum Structures and Quantum Control of Ministry of Education, School of Physics and Electronics, Hunan Normal University, Changsha, 410081, People's Republic of China.}
\author{Ali Mortezapour \thanks}
\email{mortezapour@guilan.ac.ir}
\affiliation{Department of Physics, University of Guilan, P.O. Box 41335-1914, Rasht, Iran.}
\author{Dan Long}
\author{Jia Wang}
\author{Qianqian Ma}
 \affiliation{Synergetic Innovation Center for Quantum Effects and Application, Key Laboratory of Low-dimensional Quantum Structures and Quantum Control of Ministry of Education, School of Physics and Electronics, Hunan Normal University, Changsha, 410081, People's Republic of China.}

\date{\today}
             \begin{abstract}
In this work, we study quantum evolution of an open moving-qubit modulated by a classical driving field. We obtain the density operator of qubit at zero temperature and analyze its quantum evolution dynamics by using  quantum speed limit time (QSLT) and a non-Markovianity measure introduced recently. The results show that both the non-Markovian environment and the classical driving can speed up the evolution process, this quantum speedup process is induced by the non-Markovianity and the critical points only depend on the qubit velocity. Moreover, the qubit motion will delay the evolution process, but this negative effect of the qubit velocity on the quantum speedup can be suppressed by the classical driving. Finally, we give the corresponding physical explanation by using the decoherence rates.

\end{abstract}
\maketitle

\section{Introduction}

Quantum mechanics restriction on the evolution speed of quantum systems is called quantum speed limit, which is the fundamental law of nature\cite{Lloyd.2000,Anandan.1990,Vaidman.1992,Luo.2004,H.P.BreuerandF.Petruccione.2002,Michael.A.NielsenandIsaac.L.Chuang.2010}. In the past decades, it has been attracting considerable attention and played remarkable roles in various areas of quantum physics, including quantum communication, quantum optimal control, quantum computation and non-equilibrium thermodynamics\cite{Yung.2006,Bekenstein.1981,Caneva.2009,Mukherjee.2013,Hegerfeldt.2013,Hegerfeldt.2014,Avinadav.2014,Lloyd.2002,Chatrchyan.b,Deffner.2010,Shen-Shuang Nie,Nikolai,Eoin}. Quantum speed limit time (QSLT) is defined as the minimal evolution time of a quantum system. For a unitary evolution, there are two common bounds of the QSLT. One is expressed as $\tau_{qsl}=\pi\hbar/(2\triangle E)$, where $\triangle E$ represents the energy fluctuation of the initial state, which is proposed by Mandels and Tamm(i.e. the MT bound). The other is $\tau_{qsl}=\pi\hbar/(2 E)$, where $\tau_{qsl}$ depends on the average energy $ E$, which is derived by Margolus and Levitin(i.e the ML bound). Combination of the two bounds yields, the QSLT of the two orthogonal pure states in the closed system as $\tau_{qsl}=\max\{\pi\hbar/2\triangle E,\pi\hbar/(2 E) \}$\cite{Bekenstein.1981,Fleming.1973,Bhattacharyya.1999,Vaidman.1992,Margolus.1998}. In addition, the MT-QSL (i.e. the MT-QSL bound based on the relative purity), the NI-QSL(i.e. the quantum speed limit without using the von Neumann trace inequality) and the quantum speed limit in a non-equilibrium environment as well as the QSL bound in terms of the quantum fisher information has also been investigated in succession\cite{delCampo.2013,S.X.WuY.zhangC.S.yuandH.S.Song.2014,Cai.2017,Schiro.2014,Peronaci.2015,Bhupathi.2016,OviedoCasado.2016,Lombardo.2013,Mirkin.2018,Du}. In particular, a new bound of the quantum speed limit different from the MT and the ML bounds is also proposed by employing the gauge invariant and geometric properties of quantum mechanics\cite{Yujun Zheng.2019}.

On the other hand, the measure of the non-Markovianity in dynamic processes for an open two-level system has been presented in Refs.\cite{Laine.2010,Zeng.2011,He.2017,Fanchini.2013,Benedetti.2014}. In recent years, the quantum speed limit and the non-Markovian dynamic process of an open quantum system has been widely concerned. For example, the authors in Ref.\cite{Deffner.2013} acquired the unified bound of an open system by using the Bures angle based on the ML and MT bounds and found that the non-Markovian effects could speed up the quantum evolution. The quantum speedup in open quantum systems and the relationship between the quantum speedup with the formation of a system-environment bound state is also studied in\cite{Liu H-B.2016,Mirkin N.2016,Xu K.2018,Ahansaz.2019,Wang J.2018}. In addition, quantum speedup dynamics process can be also obtained by using the coherent driving, the correlated channel and the Markovian and non-Markovian noise channels\cite{Ying-Jie Zhang.2021,N. Awasthi.2020,N. Awasthi.2022}. Recently, we have studied the QSLT and the non-Markovianity of the atom in Jaynes-Cummings model coupling with the Lorentzian reservoir and the Ohmic reservoir, in which we characterized the non-Markovianity by using the positive derivative of the trace distance, the probability of the atomic excited state and the negative decoherence rate, respectively\cite{Zou H-M.03,Zou H-M.04}.

These researches mentioned above are based on a stationary qubit model, but one is not making the atom completely stationary in the present experiments such as cavity QED and cooling technology\cite{Zhang.2012,Zhang.2014,Zhang.2014b}. Therefore, it seems logical to consider the motion of the qubits. Recently, A. Mortezapour and D. Park $et\ al$ studied the entanglement and coherence of an open moving-qubit by assuming the length of the cavity is close to infinity, and their results showed that the entanglement and coherence can be protected from decay by suitable adjusting the velocity of the qubit \cite{Mortezapour.2017,DaeKilPark.2017,AliMortezapourMahdiAhmadiBorjiDaeKilParkandRosarioLoFranco.2017,Golkar.2020}. The authors in \cite{Hong-Mei Zou.2022} investigated recently the entanglement dynamics of an open moving-biparticle system driven by classical-field and the results showed that the classical driving can not only protect the entanglement, but also effectively eliminate the influence of the qubit velocity and the detuning on the quantum entanglement. At the same time, Y. J. Zhang and W. Han studied the evolution process of an open stationary system driven by external classical field and found that the classical field can effectively speed up the evolution of the stationary qubit\cite{Zhang.2015}. Inspired by these works, we construct a model of an open moving-qubit driven by external classical field in order to understand the influence of the classical field and the velocity of moving-qubit in an infinite cavity at zero-temperature on quantum evolution process.

In fact, we try to regulate the quantum evolution process of the moving-qubit by the external classical field. Considering both the external classical field and the qubit movement will make this model more complicated, but we first obtain an analytical solution of such a qubit in the dressed-state basis when the cavity has Lorentzian spectral density by using the dressed-state. Afterward, we investigate in detail the effect of the classical field and the velocity of qubit on the QSLT. We find that both the strong qubit-cavity coupling and the classical field can accelerate the evolution process, but the velocity of moving-qubit can delay the evolution process. Namely, we provide a method to suppress the influence of the velocity of moving-qubit on the quantum evolution process by external classical field when the qubit is not completely stationary, which is the second goal of this work. In addition, we investigate a non-Markovianity measure introduced recently in the dynamic process and the relationship between the non-Markovianity and the QSLT, which is the third goal of this work.

This paper is organized as follows: In Section 2, we present the physical model and analytical solution of an open moving-qubit driven by the external classical field. In section 3, we have provided a useful preliminary to QSLT and non-Markovianity and then calculated them for the proposed system. In section 4, we give the results and discussions. Finally, a simple conclusion of this paper is given in section 5.

\section{Physical model and analytical solution}
Authors in Ref.\cite{Lang.1973} proposed a method treating an open moving-qubit model, in which the reservoir is modeled by a leaky cavity and the leakage of radiation from the cavity is replaced by the coupling to the outside world. The structure of the environment consists of two completely reflective mirrors at $z=l$ and $-L$, and a semitransparent mirror at $z=0$. It can be said that the environment consists of two consecutive cavities in intervals (L, 0) and (0, $l$). The model is shown in Fig. 1.
\begin{figure}[h!]
	\centering\includegraphics[width=8cm]{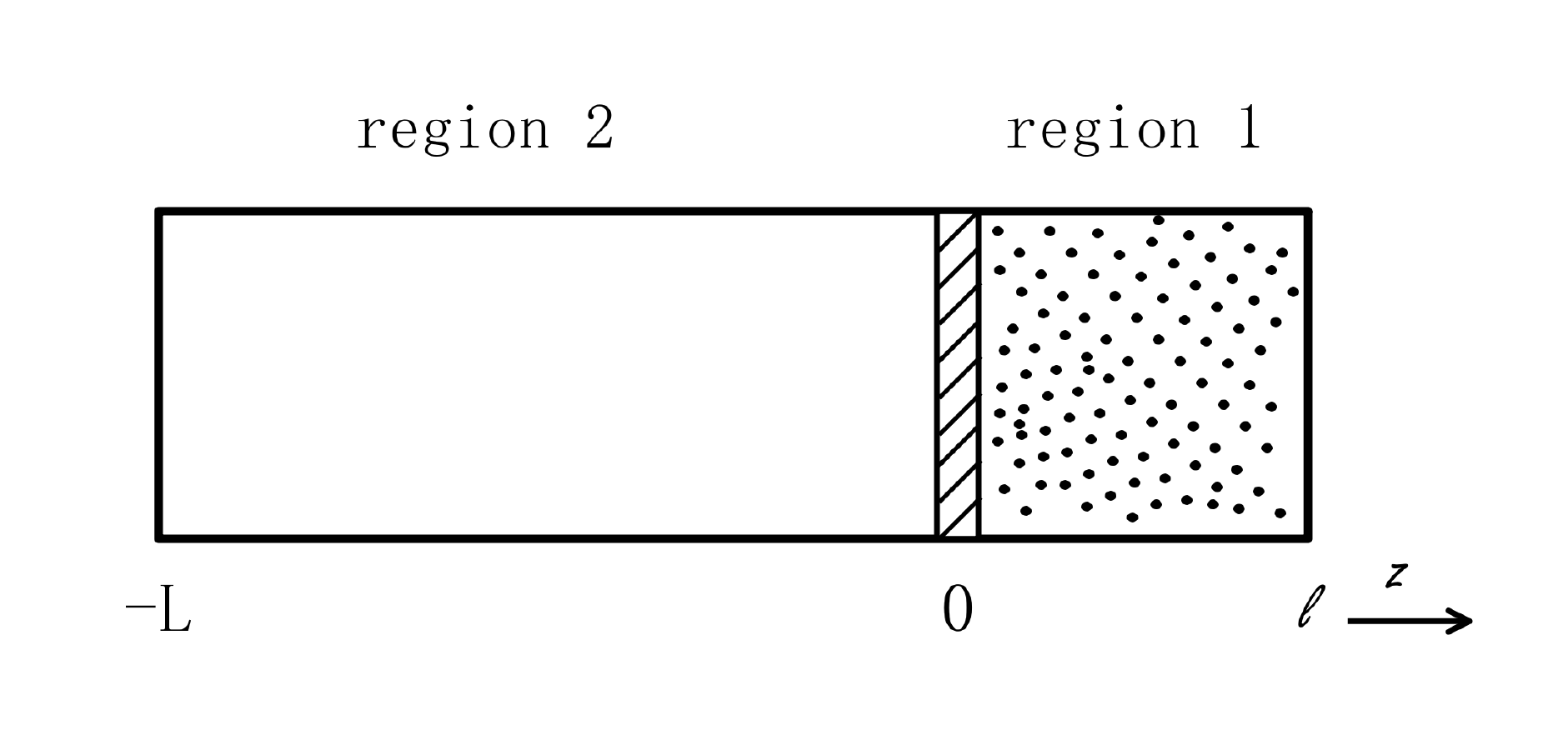}
	\caption{The reservoir is modelled by a leaky cavity with the Lorentzian spectral density. We also assume that  the reservoir is big enough, that is the cavity with partially reflecting mirror is imbedded in large ideal cavity($L\rightarrow\infty$).}
\end{figure}

Based on\cite{Lang.1973}, we structure a moving-qubit interacting with the multimode reservoir, where the qubit is driven by the classical field. In this model, the classical field is polarized along the x-axis and propagates along the y-axis, and the qubit interacts with the second cavity located in the range (0, $l$) and $l\to\infty$, the qubit also moves along the z-axis at a constant velocity $v$ as shown in Fig. 2. The Hamiltonian is given by($\hbar=1$)

\begin{equation}\label{EB01}
\begin{split}
\hat{H}=&\frac{1}{2}\omega_{0}\sigma_{z}+\sum_{k}\omega_{k}a_{k}^{\dag}a_{k}+\Omega e^{-i\omega_{f}t}\sigma_{+}+\Omega e^{i\omega_{f}t}\sigma_{-}\\
&+\sum_{k}g_{k}(f_{k}(z)a_{k}\sigma_{+}+f_{k}^{\ast}(z)a_{k}^{\dag}\sigma_{-}),
\end{split}
\end{equation}
where $\sigma_{z}=|e\rangle\langle e|-|g\rangle\langle g|, \sigma_{+}=|e\rangle\langle g|$, and $\sigma_{-}=\sigma_{+}^{\dag}$ associated with the upper level $|e\rangle$ and lower level $|g\rangle$.  $\omega_{0}$ is the transition frequency of the qubit. $a_{k}^{\dag}(a_{k})$ and $\omega_{k}$ are the creation(annihilation) operator and the frequency of the $k$-th mode of the cavity, respectively. And $\omega_{f}$ represents the frequency of the classical field. In addition, $g_{k}$ denotes the coupling constant between the qubit and the $k$-th mode of the cavity, and $\Omega$ designates the coupling strength between the qubit and the classical field. The parameter $f_{k}(z)$ describes the shape function induced by the velocity of the qubit along the $z$-axis, which it is given by

\begin{equation}\label{EB02}
\begin{split}
f_{k}(z)=f_{k}(vt)=\sin[k(z-l)]=\sin[\omega_{k}(\beta t-\tau_{0})],
\end{split}
\end{equation}
where $\beta=\upsilon/c$ and $\tau_{0}=l/c$, $\upsilon$ and $c$ are respectively the velocities of the moving-qubit and the light, $l$ is the length of the right side cavity. Note that the shape function is not zero when $z=0$, while it is zero when $z=l$(perfect boundary)\cite{AliMortezapourMahdiAhmadiBorjiDaeKilParkandRosarioLoFranco.2017}.

\begin{figure}[h!]
	\centering\includegraphics[width=8cm]{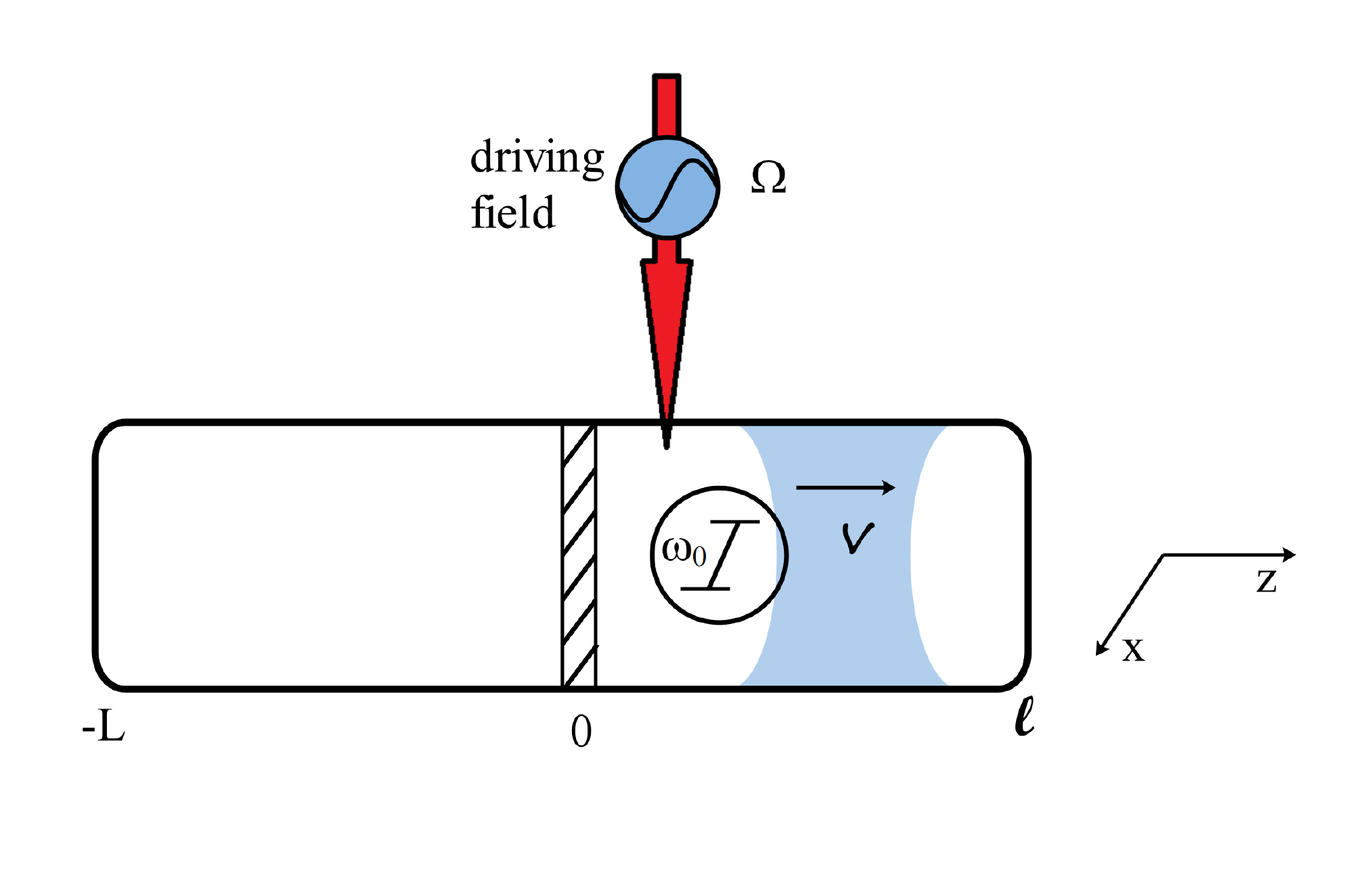}
	\caption{Schematic illustration of a setup where a single qubit is moving inside a leaky cavity and driven by the classical speedup. The qubit is a two-level atom with transition frequency $\omega_{0}$ at constant velocity $v$.}
\end{figure}

In the dressed-state basis $\{|E\rangle=\frac{1}{\sqrt{2}}(|g\rangle+|e\rangle), |G\rangle=\frac{1}{\sqrt{2}}(|g\rangle-|e\rangle)\}$ and exploiting the unitary operator $U_{1}=\exp{[-i\omega_{f}\sigma_{z}t/2]}$, the Hamiltonian in Eq. (\ref{EB01}) is equivalently transformed to an effective Hamiltonian
\begin{equation}\label{EB03}
\begin{split}
\hat{H}_{e}=&\frac{\omega_{D}}{2}\Sigma_{z}+\sum_{k}\omega_{k}a_{k}^{\dag}a_{k}\\
&+\sum_{k}g_{k}^{1}(f_{k}(z)a_{k}e^{i\omega_{f}t}\Sigma_{+}+f_{k}^{\ast}(z)a_{k}^{\dag}e^{-i\omega_{f}t}\Sigma_{-}),
\end{split}
\end{equation}
where $\omega_{D}=\sqrt{\Delta^{2}+4|\Omega|^{2}}$, $g_{k}^{1}=g_{k}/2$, $\Delta=\omega_{0}-\omega_{f}$, $\Sigma_{z}=|E\rangle\langle E|-|G\rangle\langle G|$, $\Sigma_{+}=|E\rangle\langle G|$ and $\Sigma_{-}=\Sigma_{+}^{\dag}$. For the sake of simplicity, we only discuss the reservoir at zero temperature in the following.

In the interaction picture, considering the transformation associated with the unitary operator $U_{2}=\exp\{-i[\omega_{D}\Sigma_{z}t/2+\omega_{k}a_{k}^{\dag}a_{k}t]\}$, the Hamiltonian reads
\begin{equation}\label{EB04}
\begin{split}
\hat{H}_{I}=&\sum_{k}g_{k}^{1}[f_{k}(z)a_{k}\Sigma_{+}e^{i(\omega_{D}+\omega_{f}-\omega_{k})t}\\
&+f_{k}^{\ast}(z)a_{k}^{\dag}\Sigma_{-}e^{-i(\omega_{D}+\omega_{f}-\omega_{k})t}].
\end{split}
\end{equation}
 The initial state of the total system is $|\psi(0)\rangle=C_{1}(0)|E\rangle|0\rangle_{R}$, where $|0\rangle_{R}$ denotes the vacuum state of the reservoir and $C_{1}(0)=1$. At any time $t>0$, the state of the total system is given by  $|\psi(t)\rangle=C_{1}(t)|E\rangle|0\rangle_{R}+\sum_{k}C_{k}(t)|G\rangle|1_{k}\rangle_{R}$ and meets the normalization condition $\left|\mathrm{C}_{1}(\mathrm{t})\right|^{2}+\sum_{\mathrm{k}}\left|\mathrm{C}_{\mathrm{k}}(\mathrm{t})\right|^{2}=1$, where the $|1_{k}\rangle_{R}$ is the state of the reservoir with only one excitation in the $k$-th mode. Using the Schr\"{o}dinger equation, the time evolution of the probability amplitudes are obtained as
\begin{equation}\label{EB05}
\begin{split}
&\dot{C}_{1} (t)=-i\sum_{k}g_{k}^{1}f_{k}(z)C_{k}(t)e^{i(\omega_{D}+\omega_{f}-\omega_{k}) t}\\
&\dot{C}_{k} (t)=-i g_{k}^{1}f_{k}(z) C_{1}(t)e^{-i(\omega_{D}+\omega_{f}-\omega_{k})t}.
\end{split}
\end{equation}
Since there is no excitation in the initial state of the reservoir i.e. $C_{k}(0)=0$ thus we can obtain the integral differential equation of $C_{1}(t)$ using Eq. (\ref{EB05}) can be obtained as:
\begin{equation}\label{EB06}
\begin{split}
\dot{C}_{1}(t)=-\int_{0}^{t}dt_{1}F(t-t_{1})C_{1}(t_{1}),
\end{split}
\end{equation}
the correlation function $F(t-t_{1})$ can be acquired as
\begin{equation}\label{EB07}
\begin{split}
F(t-t_{1})=&\int_{-\infty}^{\infty}J(\omega_{k}) \sin[\omega_{k} (\beta t-\tau_{0})]\times\\
&\sin[\omega_{k}(\beta t_{1}-\tau_{0})] e^{i(\omega_{D}+\omega_{f}-\omega_{k})(t-t_{1})}d\omega_{k},
\end{split}
\end{equation}
where $J(\omega_{k})$ is the spectral density of the reservoir. For the Lorentzian spectral density, $J(\omega_{k})$ adopts the following form
\begin{equation}\label{EB08}
\begin{split}
J(\omega_{k})=\frac{1}{2\pi}\frac{\gamma \lambda^{2}}{[(\omega_{0}-\omega_{k})^{2}+\lambda^{2}]},
\end{split}
\end{equation}
where $\lambda$ is the spectral width of the reservoir and $\gamma$ is the decay rate of the excited state of the qubit. The condition $\lambda>2\gamma$ and $\lambda<2\gamma$ respectively indicate the weak and strong qubit-cavity coupling regimes\cite{Breuer.2009,Zou.2014,HongMeiZou.2016}. In the weak coupling regime, the evolution of the dynamics is Markovian. But in the strong coupling regime, the evolution of the dynamics will show obvious non-Markovian characteristics.

By substituting Eq. (\ref{EB08}) into Eq. (\ref{EB07}) the correlation function $F(t-t_{1})$ can be calculated as
\begin{equation}\label{EB09}
\begin{split}
F(t-t_{1})=\sum\limits_{i=1}^{4}F_{i}(t-t_{1}),
\end{split}
\end{equation}
making use of
\begin{equation}\label{EB10}
\int_{-\infty}^{\infty} \frac{e^{-iza }}{z^{2}+\lambda^{2}} d z=\frac{\pi}{\lambda} e^{-\lambda|a|}.
\end{equation}%
The analytical solutions obtained are

\begin{equation}\label{EB11}
\begin{split}
&F_{1}\left(t, t_{1}\right)=\frac{-r \lambda}{8} e^{iw_{0}\beta \left(t+t_{1}\right)-2i w_{0}\tau_{0}+i(w_{D}-w_{0}+w_{f} )(t-t_{1})}\\&
e^{-\lambda\left |  -\beta(t+t_{1})+2 \tau_{0}+(t-t_{1}) \right |}\\
&F_{2}\left(t, t_{1}\right)=\frac{r \lambda}{8} e^{i(w_{D}-w_{0}+w_{f}+w_{0}\beta )(t-t_{1})} e^{-\lambda\left |(1-\beta)(t-t_{1})\right |}\\
&F_{3}\left(t, t_{1}\right)=\frac{r \lambda}{8} e^{i(w_{D}-w_{0}+w_{f}-w_{0}\beta )(t-t_{1})} e^{-\lambda\left |(1+\beta)(t-t_{1})  \right | }\\
&F_{4}\left(t, t_{1}\right)=\frac{-r \lambda}{8} e^{-iw_{0}\beta \left(t+t_{1}\right)+2i w_{0}\tau_{0}+i(w_{D}-w_{0}+w_{f} )(t-t_{1})}\\&
 e^{-\lambda\left |\beta(t+t_{1})-2 \tau_{0}+(t-t_{1}) \right|}
\end{split}
\end{equation}

In the continuum limit $\tau_{0}\to\infty$ and when $t> t_{1}$, $0< \beta < 1$ \cite{DaeKilPark.2017,AliMortezapourMahdiAhmadiBorjiDaeKilParkandRosarioLoFranco.2017}, the analytic solutions of Eq. (\ref{EB11}) gives rise to,

\begin{equation}\label{EB12}
\begin{split}
&F_{1}\left(t, t_{1}\right)=\frac{-r \lambda}{8} e^{iw_{0}\beta \left(t+t_{1}\right)-2i w_{0}\tau_{0}+i(w_{D}-w_{0}+w_{f} )(t-t_{1})}\\&
e^{-\lambda[-\beta(t+t_{1})+2 \tau_{0}+(t-t_{1})]}\\
&F_{2}\left(t, t_{1}\right)=\frac{r \lambda}{8} e^{i(w_{D}-w_{0}+w_{f}+w_{0}\beta )(t-t_{1})} e^{-\lambda[(1-\beta)(t-t_{1})]}\\
&F_{3}\left(t, t_{1}\right)=\frac{r \lambda}{8} e^{i(w_{D}-w_{0}+w_{f}-w_{0}\beta )(t-t_{1})} e^{-\lambda[(1+\beta)(t-t_{1})]}\\
&F_{4}\left(t, t_{1}\right)=\frac{-r \lambda}{8} e^{-iw_{0}\beta \left(t+t_{1}\right)+2i w_{0}\tau_{0}+i(w_{D}-w_{0}+w_{f} )(t-t_{1})}\\&
 e^{-\lambda[-\beta(t+t_{1})+2 \tau_{0}-(t-t_{1})]}
\end{split}
\end{equation}

The case that the length of cavity is infinite, i.e. $\tau_{0}\to\infty$, so $F_{1}\left(t, t_{1}\right)\to 0$, $F_{4}\left(t, t_{1}\right)\to 0$. Eq. (\ref{EB09}) can be simplified as
\begin{equation}\label{EB13}
\begin{split}
F\left(t, t_{1}\right)=\frac{r \lambda}{8}\left[e^{(\mu \beta-\eta)(t-t_{1})}+e^{-(\mu \beta+\eta)(t-t_{1})}\right]
\end{split}
\end{equation}
where $\mu=\lambda+i\omega_{0}$ and $\eta=\lambda+i(\omega_{0}-\omega_{D}-\omega_{f})$.
inserting Eq. (\ref{EB13}) into Eq. (\ref{EB06}) and then using the Laplace transform, we can get $C_{1}(t)$,

\begin{equation}\label{EB14}
\begin{split}
C_{1}(t)=&\frac{\left(s_{1}-\varepsilon_{0}\right)\left(s_{1}-\varepsilon_{1}\right)}{\left(s_{1}-s_{2}\right)\left(s_{1}-s_{3}\right)} e^{s_{1}t}+\\&
\frac{\left(s_{2}-\varepsilon_{0}\right)\left(s_{2}-\varepsilon_{1}\right)}{\left(s_{2}-s_{1}\right)\left(s_{2}-s_{3}\right)} e^{s_{2}t}+
\\&\frac{\left(s_{3}-\varepsilon_{0}\right)\left(s_{3}-\varepsilon_{1}\right)}{\left(s_{3}-s_{1}\right)\left(s_{3}-s_{2}\right)} e^{s_{3}t}
\end{split}
\end{equation}
where $s_{k}(k=1,2,3)$ are the roots of the equation $s^{3}-(\varepsilon_{0}+\varepsilon_{1})s^{2}+(\varepsilon_{0}\varepsilon_{1}+\frac{\gamma \lambda }{4} )s-\frac{\gamma \lambda(\varepsilon_{0}+\varepsilon_{1}) }{8}=0$, among $\varepsilon_{0}=\mu\beta-\eta$ and $\varepsilon_{1}=-\eta-\mu\beta$. the density matrix of the qubit in the basis$\{|E\rangle, |G\rangle\}$ at time $t$ is given by
\begin{equation}\label{EB15}
\begin{split}
\rho(t)=\left(
\begin{array}{cc}
\rho_{EE}(0)|C_{1}(t)|^{2}&\rho_{EG}(0)C_{1}(t)\\
\rho_{GE}(0)C_{1}^{*}(t)&1-\rho_{EE}(0)|C_{1}(t)|^{2}
\end{array}
\right).
\end{split}
\end{equation}

We can write the master equation for the density operator of the qubit, i.e.
\begin{equation}\label{EB16}
\begin{split}
\dot{\rho}(t)=&\mathcal{L}\rho(t)\\
=&-\frac{i}{2}S(t)[\Sigma_{+}\Sigma_{-},\rho(t)]\\
&+\Gamma(t))\{\Sigma_{-}\rho(t)\Sigma_{+}-\frac{1}{2}\Sigma_{+}\Sigma_{-}\rho(t)-\frac{1}{2}\rho(t)\Sigma_{+}\Sigma_{-}\}.
\end{split}
\end{equation}
where $S(t)$ and $\Gamma(t)$ are respectively time-dependent Lamb shift and the decoherence rate of the qubit.
\begin{equation}\label{EB17}
\begin{split}
S(t)=-2\Im \{\frac{\dot{C}_{1}(t)}{C_{1}(t)}\},\\
\Gamma(t)=-2\Re \{\frac{\dot{C}_{1}(t)}{C_{1}(t)}\},
\end{split}
\end{equation}

\section{Quantum speed limit and non-Markovianity}
In this section, we briefly review the QSLT and the non-Markovianity for an open quantum system. The bound of the minimal evolution time from an initial state $\rho_{0}$ to a final state $\rho(\tau)$ is defined as the quantum speed limit time (QSLT) of a system, where $\tau$ is an standard evolution time. As a measure of statistical distance between quantum states, the Bures angle was simplified as $B(\rho_{0},\rho(\tau))=\arccos[\sqrt{\langle\psi_{0}|\rho(\tau)|\psi_{0}\rangle}]$ in open quantum systems when $\rho_{0}=|\psi_{0}\rangle\langle\psi_{0}|$\cite{Deffner.2013}.

Based on the von Neumann trace inequality and the Cauchy-Schwarz inequality, the QSLT is obtained as follows\cite{Deffner.2013}, and it describes the bounds on all possible evolution times of an open quantum system from its initial state to its final state.
\begin{equation}\label{EB18}
\begin{split}
\tau_{qsl}=\max\{\frac{1}{\mathcal{V}_{\tau}^{op}},\frac{1}{\mathcal{V}_{\tau}^{tr}},\frac{1}{\mathcal{V}_{\tau}^{hs}}\}\sin^{2}[B(\rho_{0},\rho(\tau))],
\end{split}
\end{equation}
where
$\mathcal{V}_{\tau}^{op}=\frac{1}{\tau}\int_{0}^{\tau}dt||\mathcal{L}\rho(t)||_{op}$, $\mathcal{L}\rho(t)$ is the time-dependent non-unitary dynamical operator, and $||.||_{op}$ represents operator norm. $\mathcal{V}_{\tau}^{tr}=\frac{1}{\tau}\int_{0}^{\tau}dt||\mathcal{L}\rho(t)||_{tr}$, $||.||_{tr}$ indicates trace norm. $\mathcal{V}_{\tau}^{hs}=\frac{1}{\tau}\int_{0}^{\tau}dt||\mathcal{L}\rho(t)||_{hs}$ and $||.||_{hs}$ shows Hilbert-Schmidt norm. Owning to the relationship $\frac{1}{\mathcal{V}_{\tau}^{op}} >\frac{1}{\mathcal{V}_{\tau}^{hs}} > \frac{1}{\mathcal{V}_{\tau}^{tr}}$, it is easy to prove that the ML bound based on the operator norm provides the sharpest bound of quantum speed limit time of open quantum
systems\cite{Deffner.2013}.

Regarding the Eq. (\ref{EB14}) and Eq. (\ref{EB15}), and supposing $\rho_{0}=|E><E|$, the QSLT is obtained as \cite{Xu.2014}
\begin{equation}\label{EB19}
\begin{split}
 \frac{\tau_{qsl}}{\tau} =\frac{1-|C_{1}(\tau)|^2}{\int_{0}^{\tau}|\partial_{t}|C_{1}(t)|^2|dt},
\end{split}
\end{equation}
In Eq. (\ref{EB19}) $\tau_{qsl}$ is the minimal evolution time, $\tau$ is the standard evolution time. $\frac{\tau_{qsl}}{\tau}=1$ represents that the minimal evolution time is equal to the standard evolution time, which is called as the no-speedup evolution process. $\frac{\tau_{qsl}}{\tau}<1$ indicates that the minimal evolution time is less than the standard evolution time, which is called as the speedup evolution process. And the smaller value of $\frac{\tau_{qsl}}{\tau}$ will be corresponding to the greater capacity for potential promotion of the evolution speed.

The non-Markovianity measure($\mathcal{N}$) is defined as\cite{Liu H-B.2016,Xu.2014}
\begin{equation}\label{EB20}
\begin{split}
\mathcal{N} = \max\limits_{\rho_{1}(0),\rho_{2}(0)}\int_{\sigma>0}\sigma[t,\rho_{1}(0),\rho_{2}(0)]dt,
\end{split}
\end{equation}
where $\sigma[t,\rho_{1}(0),\rho_{2}(0)]=\dot{D}[\rho_{1}(t),\rho_{2}(t)]$ is the time change rate of the trace distance. $D[\rho_{1}(t),\rho_{2}(t)]=\frac{1}{2} \text{Tr}||\rho_{1}(t)-\rho_{2}(t)||$ and indicates the distinguish ability between the two states $\rho_{1,2}(t)$ evolving from their respective initial forms $\rho_{1,2}(0)$.
$\sigma[t,\rho_{1}(0),\rho_{2}(0)]<0$, corresponds to all dynamical semigroups and all time-dependent
Markovian processes, a process is non-Markovian if there exists a pair of initial states and at certain time t such that $\sigma[t,\rho_{1}(0),\rho_{2}(0)]> 0$. We should take the maximum over all initial states $\rho_{1,2}(0)$ to calculate the degree of non-Markovianity. Similar to Refs.\cite{Zhang.2015}, by drawing a sufficiently large sample of random pairs of initial states, the optimal state pair is attained for the initial states $\rho_{1}(0)=|E\rangle\langle E|$ and $\rho_{2}(0)=|G\rangle\langle G|$ in the dressed-state basis. For the Eq. (\ref{EB15}), it can been proven that the optimal pair of initial states to maximize $\mathcal{N}$ are $\rho_{1}(0)=|E\rangle\langle E|$ and $\rho_{2}(0)=|G\rangle\langle G|$. The trace distance between of the evolved states can be written as $D[\rho_{1}(t),\rho_{2}(t)]=|C_{1}(t)|^{2}$. Thus the $\mathcal{N}$ in Eq. (\ref{EB20}) can be rewritten as
\begin{equation}\label{EB21}
\begin{split}
\mathcal{N} = \frac{1}{2}[\int_{0}^{\tau}|\partial_{t} |C_{1}(t)|^{2}|dt+|C_{1}(\tau)|^{2}-1],
\end{split}
\end{equation}
Form Eq. (\ref{EB19}) and Eq. (\ref{EB21}), the relationship\cite{Xu.2014} between the QSLT and the non-Markovianity can be obtained as
\begin{equation}\label{EB22}
\begin{split}
\frac{\tau_{qsl}}{\tau} = \frac{1-|C_{1}(\tau)|^{2}}{1-|C_{1}(\tau)|^{2}+2\mathcal{N}}.
\end{split}
\end{equation}
Eq. (\ref{EB22}) shows that the QSLT is equal to the standard evolution time when $\mathcal{N}=0$, but the QSLT is smaller than the standard evolution time when $\mathcal{N}>0$. That is, the non-Markovianity in the dynamics process can lead to the faster quantum evolution and the smaller QSLT.

\section{Results and discussion}
In this section, we will use the QSLT and the non-Markovianty to research the quantum evolution dynamics of open moving-qubit modulated by a classical driving field in detail.
\begin{figure}[h!]
	\centering\includegraphics[width=8cm]{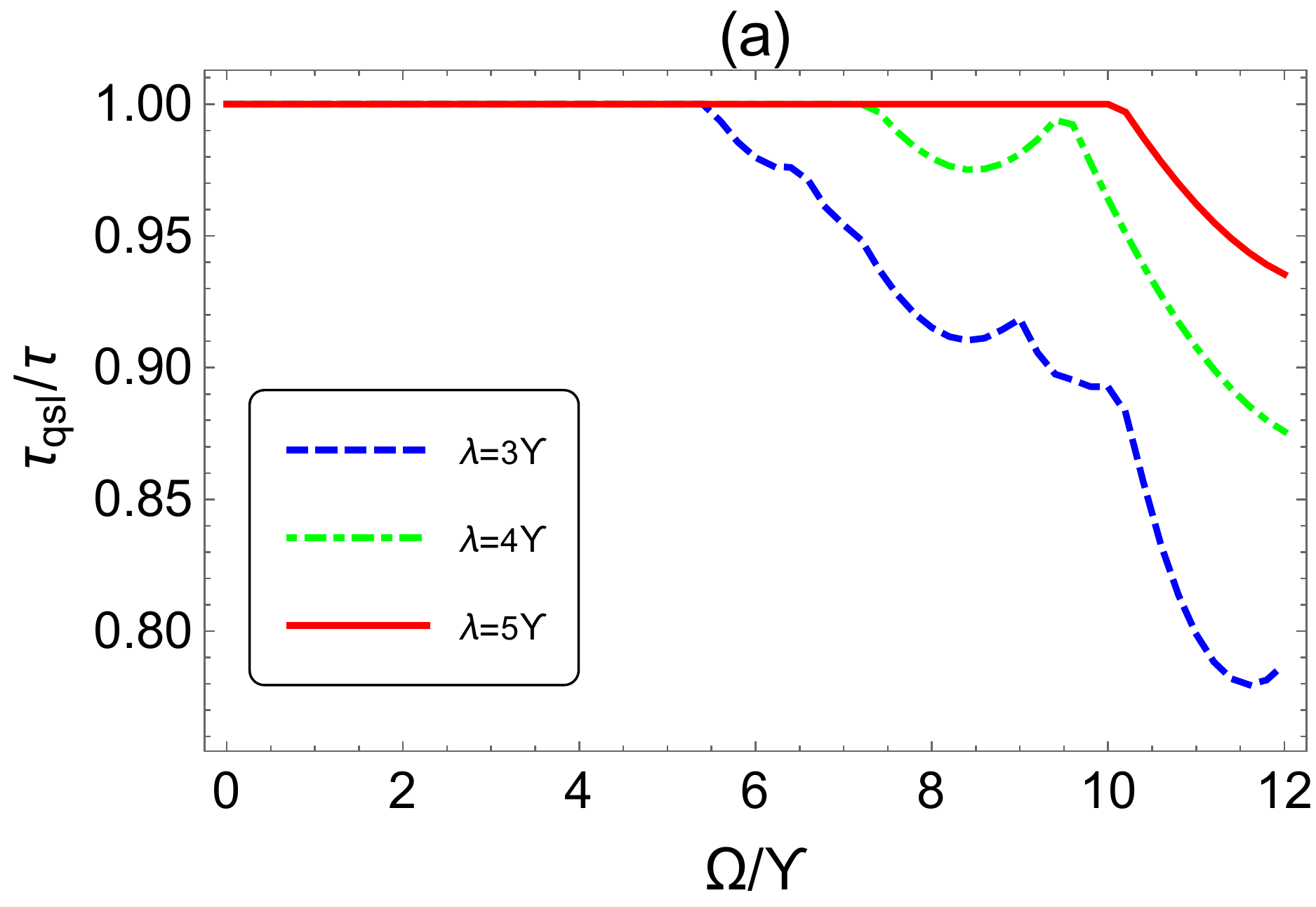}
	\centering\includegraphics[width=8cm]{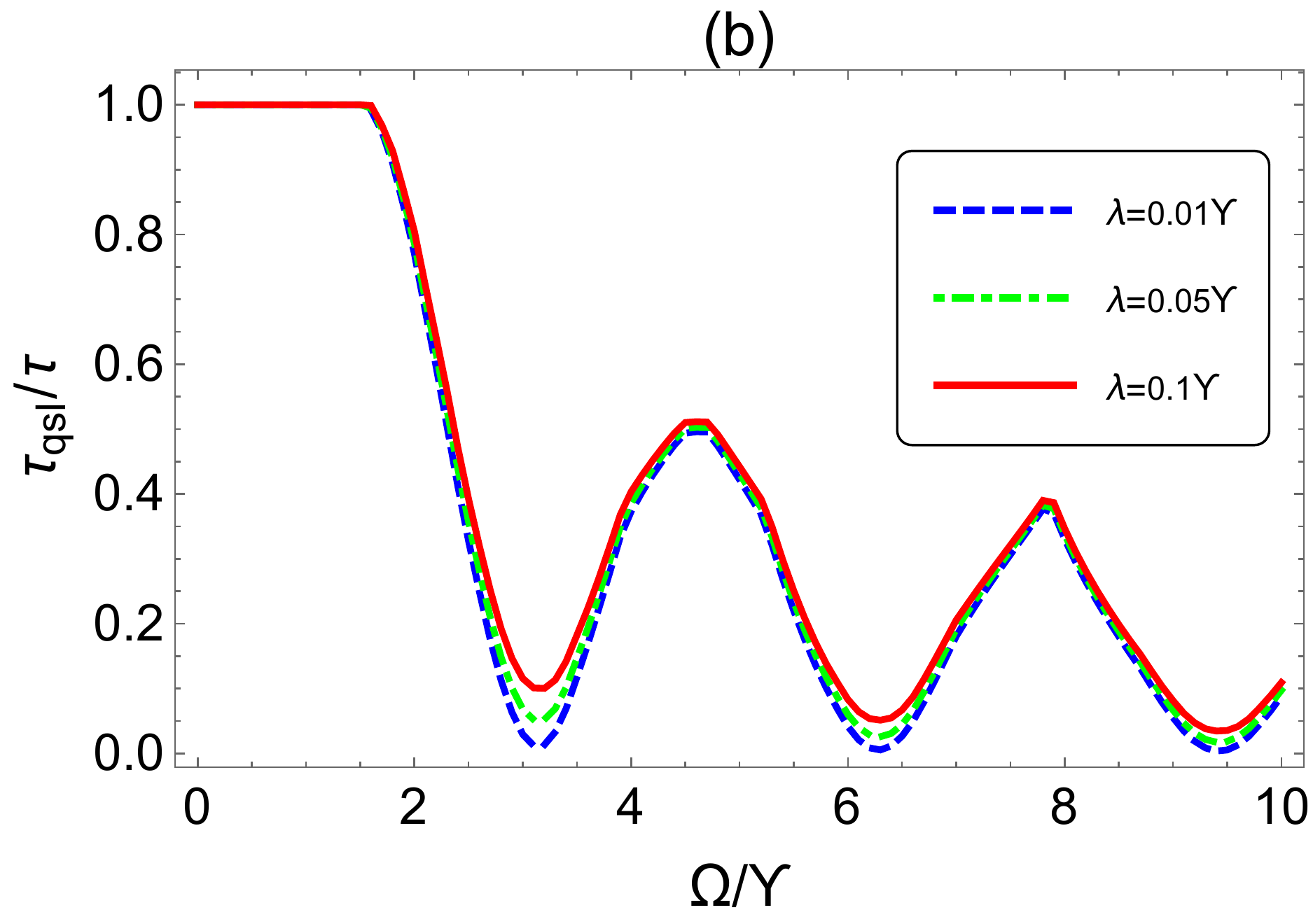}
	\caption{$\frac{\tau_{qsl}}{\tau}$ for an open system driven by an external classical field as a function of the parameter variable $\Omega/\gamma$. (a) in the weak-coupling regime($\lambda>2\gamma$). (b) in the strong-coupling regime($\lambda<2\gamma$). The velocity ratio $\beta=0$. The transition frequency $\omega_{0}=1.53\times10^{9}$. The coupling strength $\gamma=1$. The actual evolution time $\tau=1$. The detuning $\Delta=0$.}
\end{figure}

In order to obtain the quantum-accelerated evolution process manipulated by classical field under weak and strong coupling regimes, we draw Fig. 3. Fig. 3 exhibits the curves of the $\tau_{qsl}$ versus the driving strength $\Omega$ when $\beta=0$ in the weak and strong coupling regimes, respectively. From Fig. 3(a), we find that the $\tau_{qsl}$ will be equal to the standard evolution time $\tau$ when the classical driving strength $\Omega$ is small. The $\tau_{qsl}$ is less than the standard evolution time only when the classical driving strength is greater than a certain critical value in which a sudden transition from standard evolution process to speedup process will occur. For different spectral width $\lambda$, the classical driving has different critical values $\Omega_{c}$ and the speedup processes have also obvious differences. The smaller $\lambda$ is, the smaller the critical driving strength $\Omega_{c}$ is, the smaller the $\tau_{qsl}$ is, the faster the qubit evolves. This fact indicates that, in the weak coupling regime, the larger driving strength can speed up the evolution process and the quantum evolution process relies mainly on the classical driving. Fig. 3(b) shows that, in the strong coupling regime, the smaller driving strength can also speed up the evolution process and the $\tau_{qsl}$ will reduces rapidly and then oscillate due to the flowback information of the non-Markovian environment when the driving strength increases. For different spectral width $\lambda$, there is the same critical driving strength $\Omega_{c}$ and the oscillation periods of the curves are the same, but the evolution curves of speedup process are not very different. This indicates that, in the strong coupling regime, both of the classical driving and the non-Markovian environment can speed up the quantum evolution. Namely, the quantum evolution process is determined by both the classical driving and the non-Markovian environment. One important point to be noted is that, if there is not the classical driving, the quantum evolution of an open system would never be accelerated in the weak coupling regime, because the information flows irreversibly from the system to the reservoir under the dissipation of Markovian reservoir. In contrast, the quantum evolution can be accelerated in the strong coupling regime, because the information can feed back into the system from the reservoir under the memory and feedback of non-Markovian reservoir. However, if there exists the classical driving, the quantum evolution of an open system will be also accelerated in the weak coupling regime, this shows that the classical driving can be equivalent to a non-Markovian environment of the open system.

\begin{figure}[h!]
	\centering\includegraphics[width=8cm]{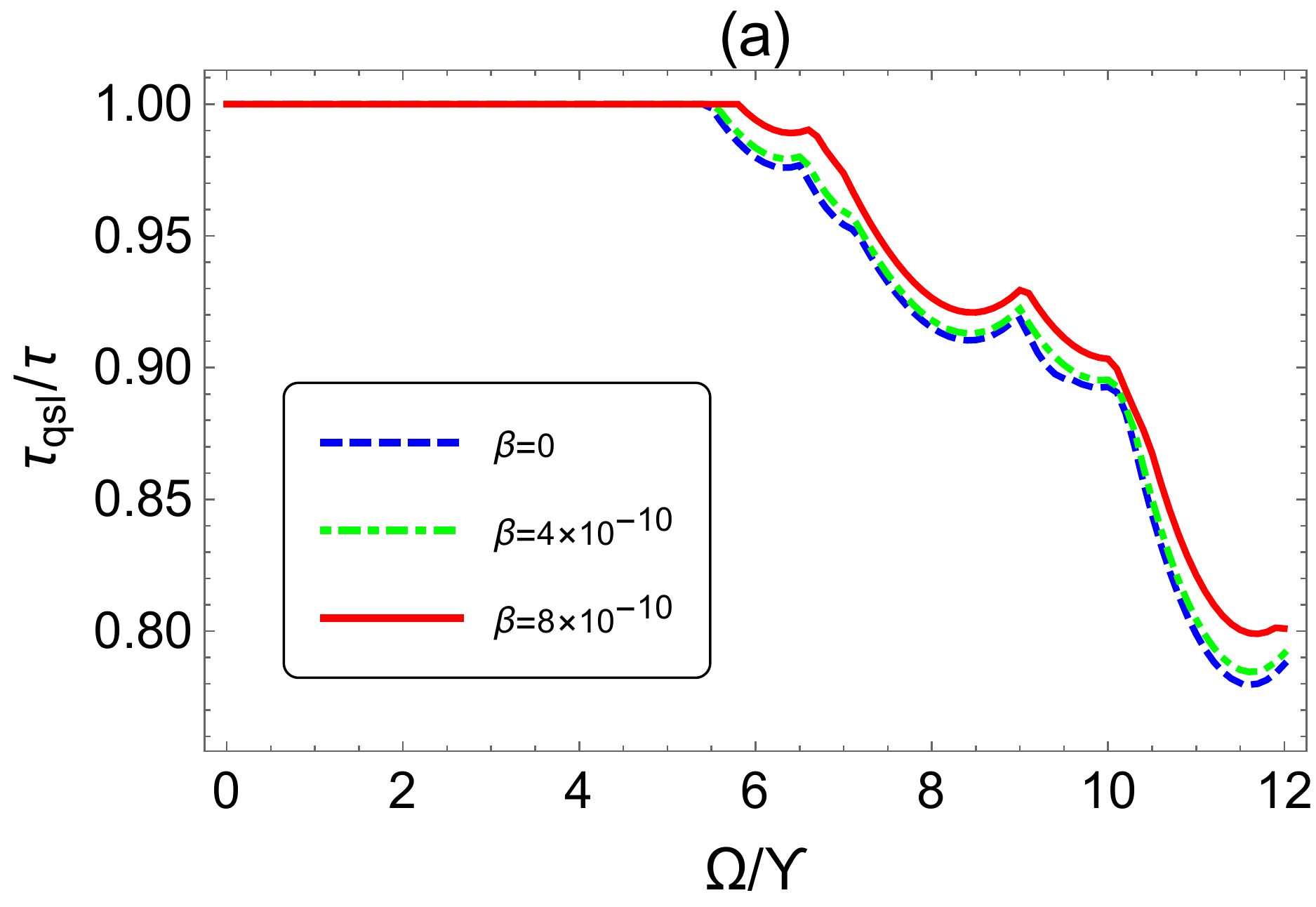}
	\centering\includegraphics[width=8cm]{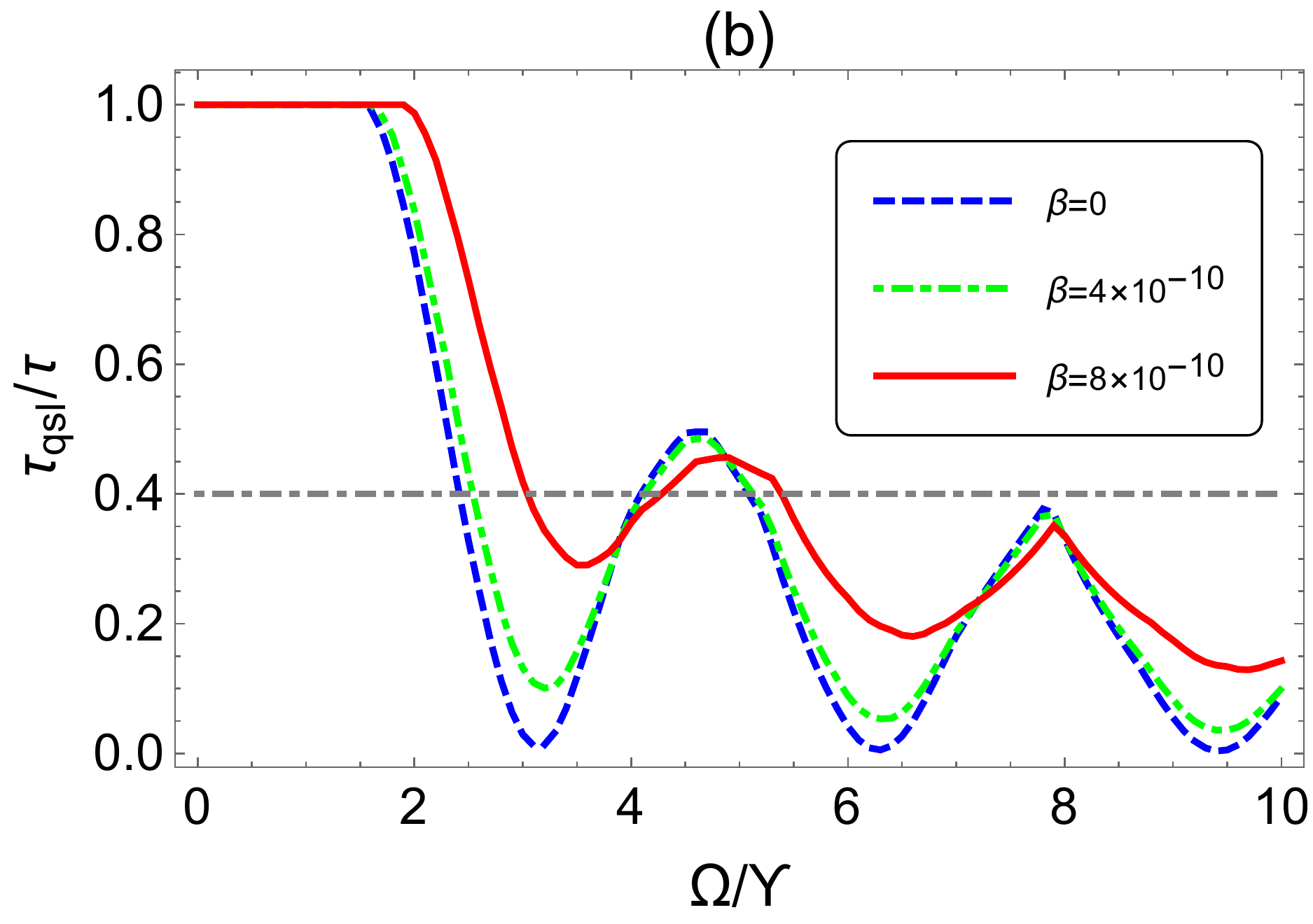}
	\caption{$\frac{\tau_{qsl}}{\tau}$ for an open system driven by an external classical field as a function of the parameter variable $\Omega/\gamma$.  (a) in the weak-coupling regime($\lambda=3\gamma$). (b) in the strong-coupling regime($\lambda=0.01\gamma$). The transition frequency $\omega_{0}=1.53\times10^{9}$. The coupling strength $\gamma=1$. The actual evolution time $\tau=1$. The detuning $\Delta=0$.}
\end{figure}

Fig. 4 shows the effect of velocity of moving-qubit and the classical driving on the $\tau_{qsl}$ in the weak and strong coupling regimes. From Fig. 4(a), we know that the evolution process of moving-qubit will be speed up in the weak-coupling regime($\lambda=3\gamma$) when $\Omega>\Omega_{c}$. The critical value $\Omega_{c}$ is dependent on the velocity ratio $\beta$ and the larger velocity ratio $\beta$ corresponds to the larger critical value $\Omega_{c}$ and the slower evolution process. That is to say, the quantum evolution process depends on both the classical driving and the velocity of moving-qubit. Comparing with Fig. 4(a), the difference of Fig. 4(b) is that the effect of velocity on the quantum evolution process is more obvious and tangible in the strong-coupling regime($\lambda=0.01\gamma$) than that in the weak-coupling regime($\lambda=3\gamma$). And, the $\tau_{qsl}$ will reduce rapidly and then oscillate due to the flowback information of the non-Markovian environment when the driving strength increases. Therefore, the velocity of moving-qubit will delay the evolution process while the classical driving can speed up the evolution process under both weak and strong coupling regimes. Namely, when the qubit moves with different velocities, the qubit can also evolve at the same speed by adjusting the classical driving strength, shown as the points at which the gray dotted line intersects these curves of $\tau_{qsl}/\tau$ in Fig. 4(b). The results show that the classical field strength can be served as a controlling tool to regulate the effect of the velocity of the moving qubit on the quantum evolution process when the atom cannot be completely stationary in the experiments of cavity QED and cooling technology.

\begin{figure}[h!]
	\centering\includegraphics[width=8cm]{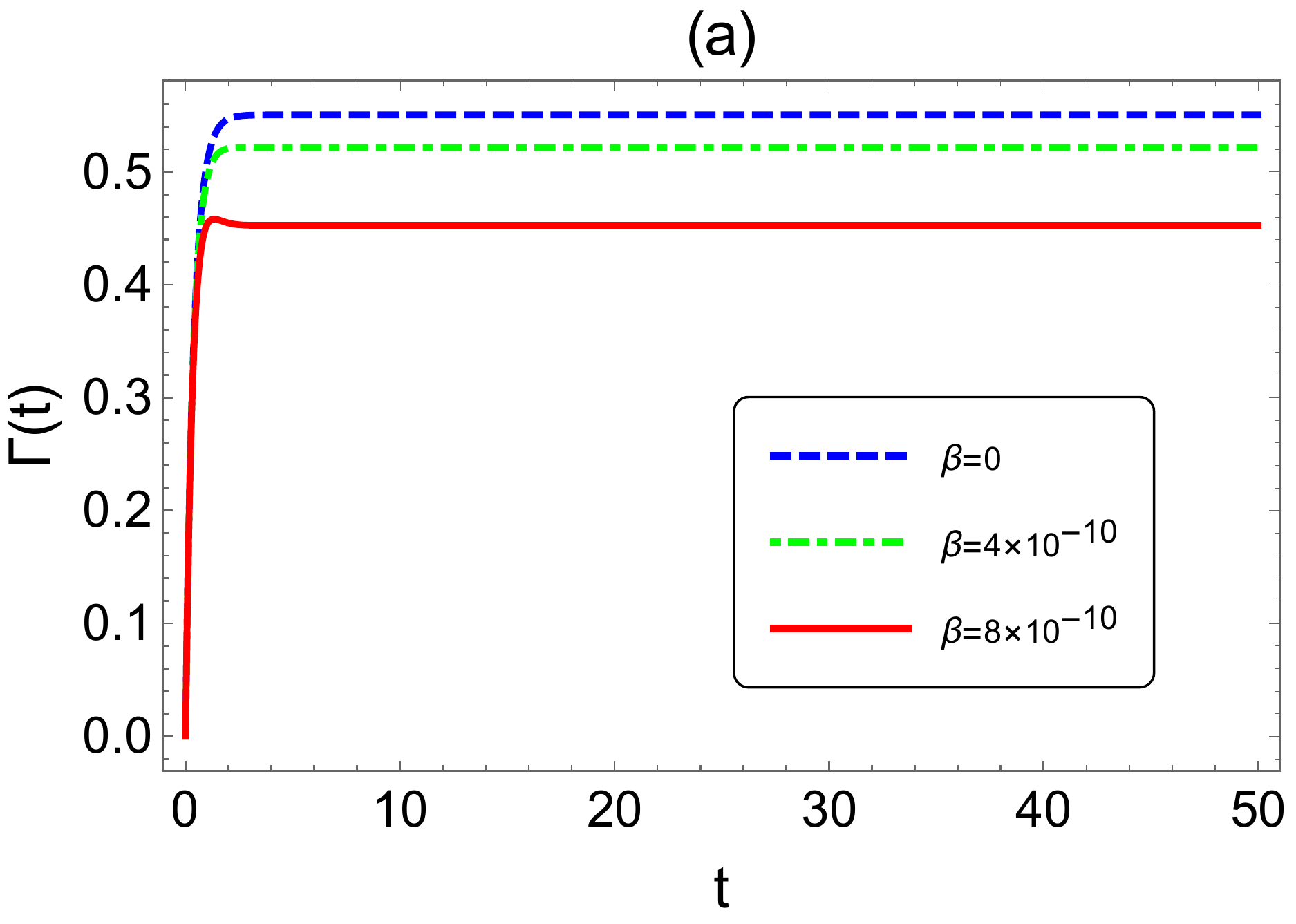}
	\centering\includegraphics[width=8cm]{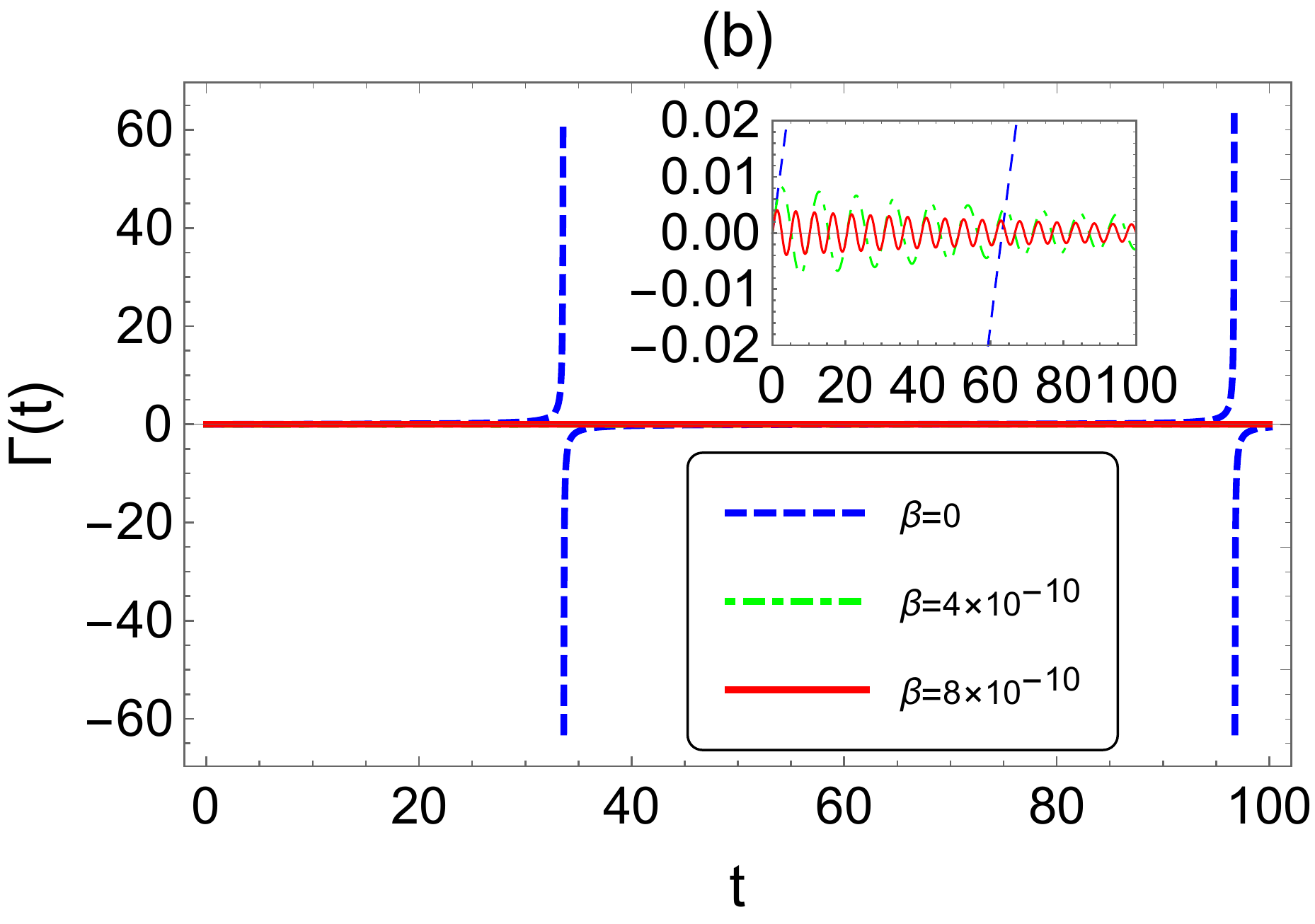}
	\caption{The decoherence rate $\Gamma(t)$ as function of the time $t$. (a) in the weak-coupling regime($\lambda=3\gamma$) and (b) in the strong-coupling regime($\lambda=0.01\gamma$). The driving strength $\Omega=0$. The transition frequency $\omega_{0}=1.53\times10^{9}$. The coupling strength $\gamma=1$. The detuning $\Delta=0$.}
\end{figure}

In order to explore the physical mechanism of the influence of the velocity of moving-qubit on the $\tau_{qsl}$, we draw the curve of the decoherence rate $\Gamma(t)$ in Fig. 5. As it is abundantly clear, the information and energy are exchanged between the qubit and the cavity, $\Gamma(t)>0$ indicates that the information flows irreversibly from the qubit to the environment, but $\Gamma(t)<0$ shows that the information flows back from the environment to the qubit. It is seen that in the weak-coupling regime($\lambda=3\gamma$) in Fig. 5(a), the decoherence rate $\Gamma(t)$ is always greater than zero, which implies that the information flows to environment without flowback. The greater the velocity of moving-qubit, the smaller the decoherence rate and, consequently, the slower the qubit evolution. Note that, the impact of the velocity on the decoherence rate is minor and then there is no flowback information that affects the moving-qubit, which is corresponding to those monotonous decay curves of the $\tau_{qsl}$ in Fig. 4(a). However, the decoherence rate $\Gamma(t)$ in Fig. 5(b) is negative in some moments, which signifies that the information flows back to the qubit from the environment in the strong-coupling regime($\lambda=0.01\gamma$). The faster the qubit, the smaller the decoherence rate, the slower the qubit evolves. In particular, the decoherence rate is heavily dependent on the velocity in the strong-coupling regime owing to the flowback information from the environment which affects the moving-qubit, such a behavior reveals the physical mechanism behind the curves in Fig. 4(b).

\begin{figure}[h!]
	\centering\includegraphics[width=8cm]{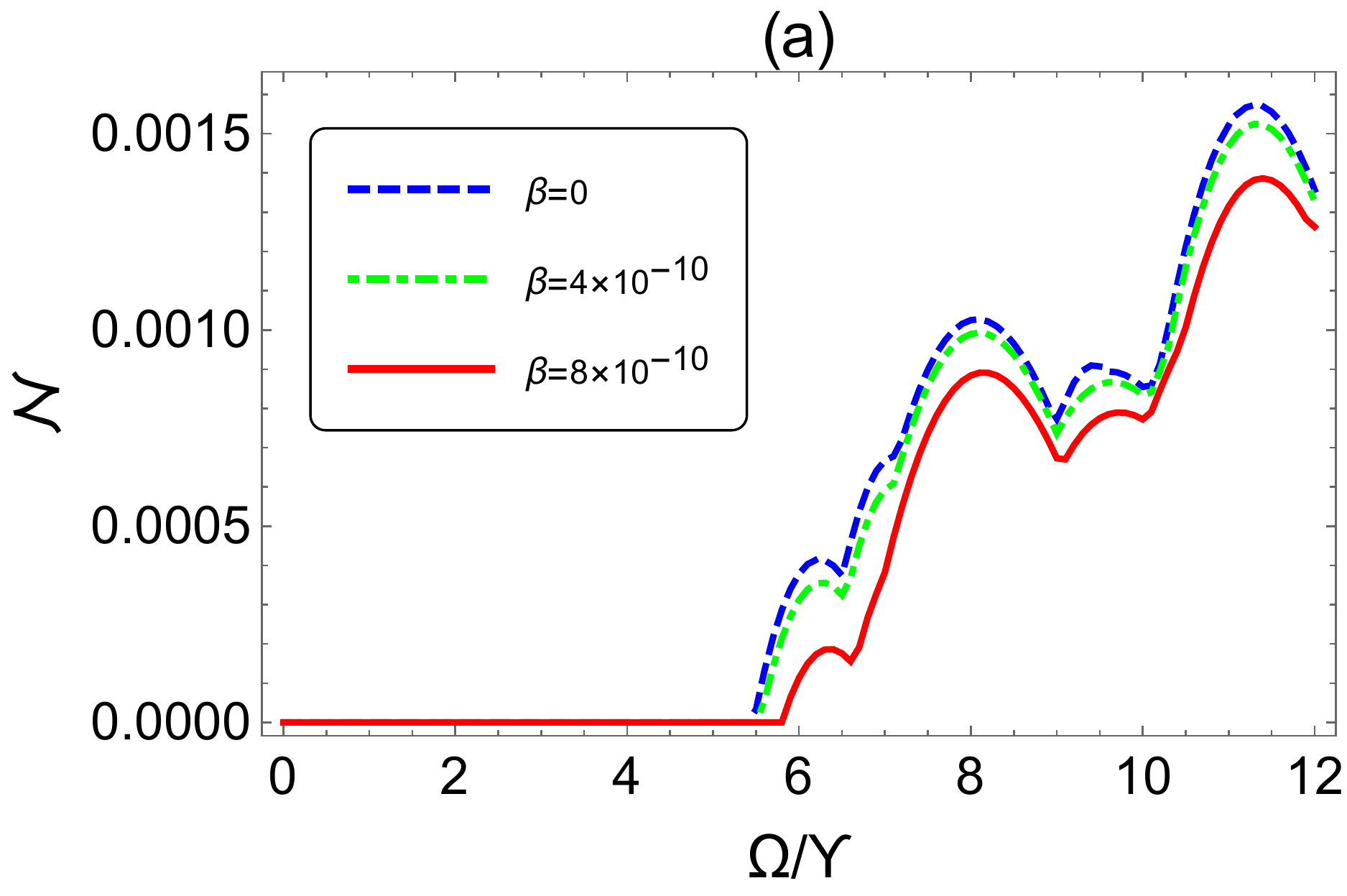}
	\centering\includegraphics[width=8cm]{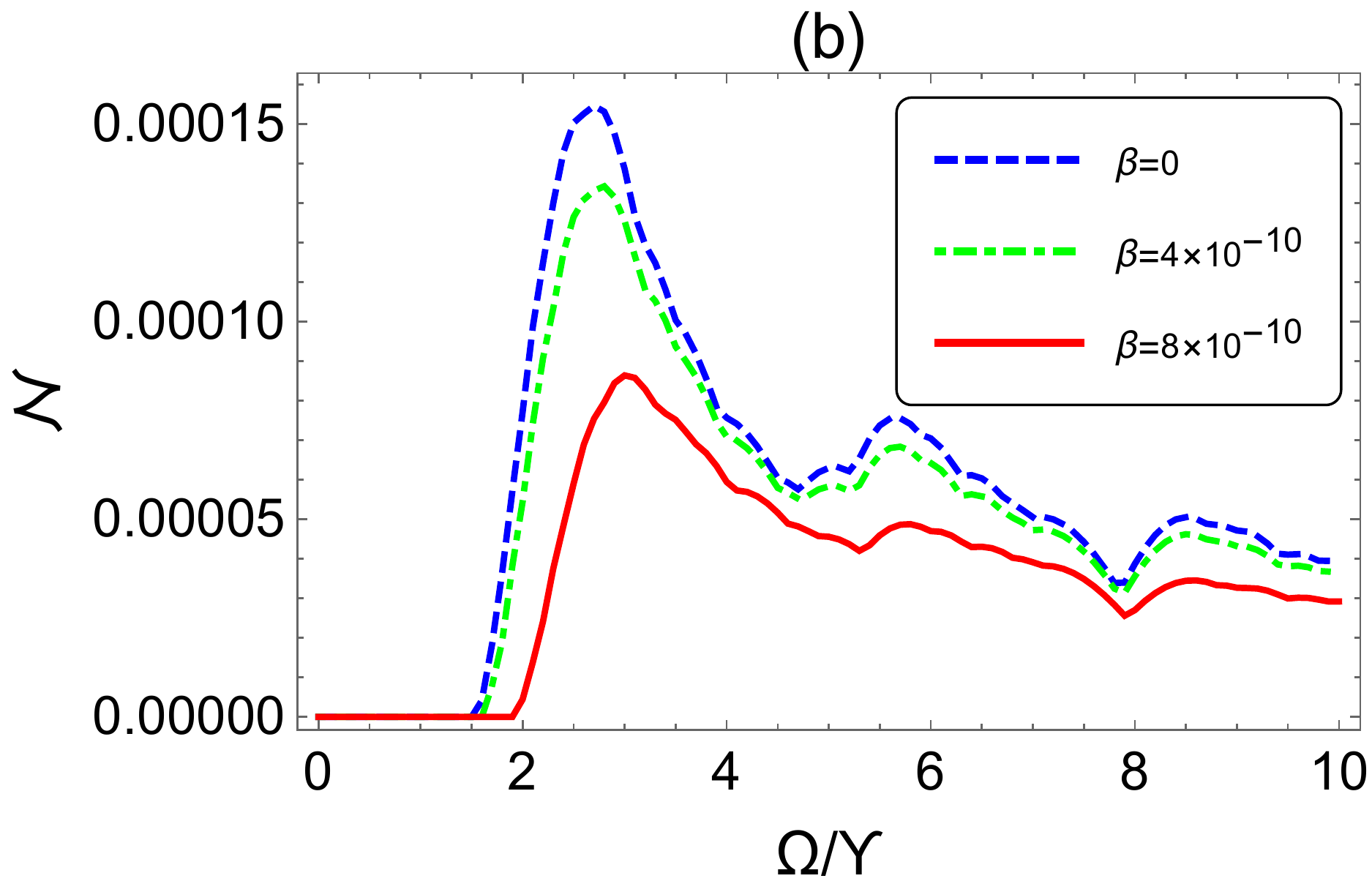}
	\caption{The non-Markovianity $\mathcal{N}$ for an open system driven by an external classical field as a function of the parameter variable $\Omega/\gamma$. (a) in the weak-coupling regime($\lambda=3\gamma$) and (b) in the strong-coupling regime($\lambda=0.01\gamma$). The transition frequency $\omega_{0}=1.53\times10^{9}$. The coupling strength $\gamma=1$. The actual evolution time $\tau=1$. The detuning $\Delta=0$.}
\end{figure}

Fig. 6 investigates the effects of the driving classical field on the non-Markovianity $\mathcal{N}$ when qubits have different velocities in the weak and strong coupling regimes. That a suitable classical driving strength can transform the original Markovian dynamics into non-Markovian dynamics. Moreover, as a common result in both weak and strong regimes, increasing the velocity of qubit leads to the critical driving strength increase. Furthermore, it is observed that the larger the value of qubit velocity, the smaller the non-Markovianity. This implies that moving qubit can significantly interfere with the backflow of information from the environment to qubits. And, the non-Markovianity increases non-monotonically when $\Omega>\Omega_{c}$.

\begin{figure}[h!]
	\centering\includegraphics[width=8cm]{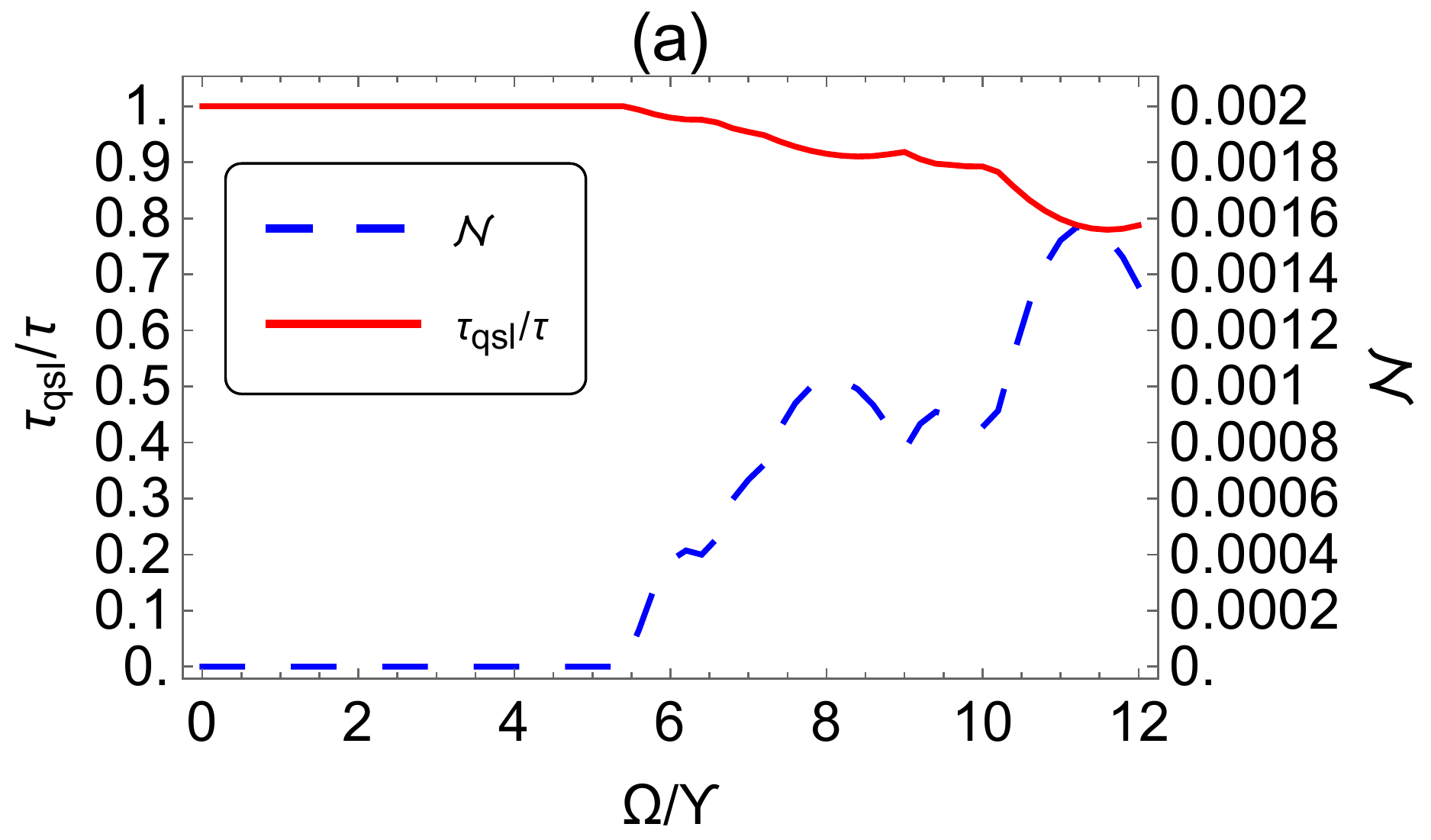}
	\centering\includegraphics[width=8cm]{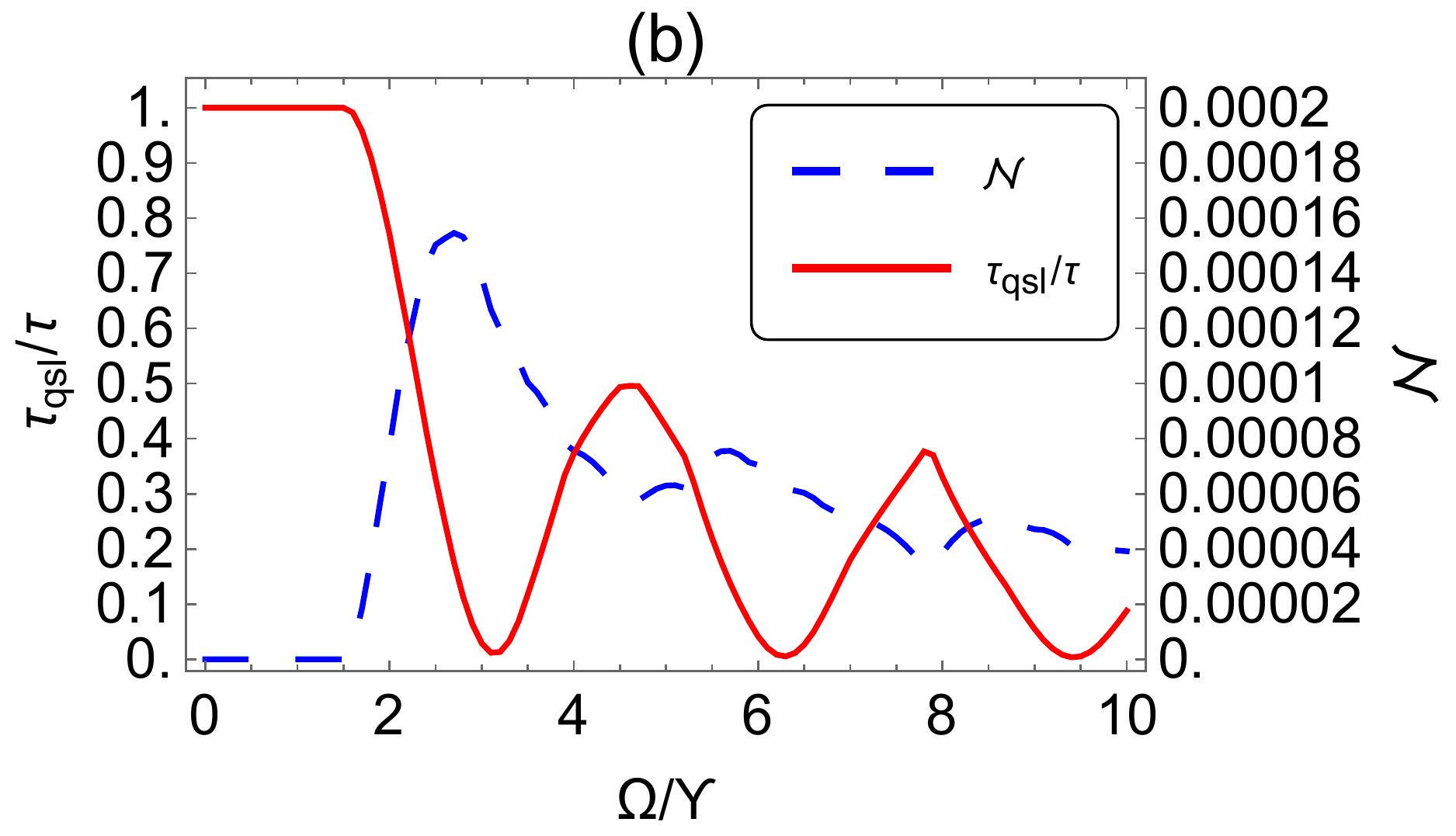}
	\caption{$\frac{\tau_{qsl}}{\tau}$ and the non-Markovianity $\mathcal{N}$ for an open system driven by an external classical field as a function of the parameter variable $\Omega/\gamma$. (a) in the weak-coupling regime ($\lambda=3\gamma$) and (b) in the strong-coupling regime($\lambda=0.01\gamma$). The coupling strength $\gamma=1$. The velocity ratio $\beta=0$. The transition frequency $\omega_{0}=1.53\times10^{9}$. The actual evolution time $\tau=1$. The detuning $\Delta=0$.}
\end{figure}

In order to shed light on the dependency relationship of the $\tau_{qsl}$ and the non-Markovianity $\mathcal{N}$, we draw Fig. 7. Fig. 7 simultaneously depicts the $\tau_{qsl}$ and the non-Markovianity $\mathcal{N}$ with respect to the driving strength $\Omega$ when $\beta=0$ in the both weak and strong coupling regimes. From Fig. 7(a), it is observed that when $\tau_{qsl}$ is equal to 1, the $\mathcal{N}$ is equal to 0. As the $\mathcal{N}$ increases, the $\tau_{qsl}$ decreases, which means that the evolution of the qubit is speeding up. Moreover Fig. 7(b) demonstrates that, in the strong-coupling regime, the curves of $\tau_{qsl}$ and $\mathcal{N}$ show oscillating behaviors due to the feedback and memory of environment. Obtaining such results are rooted in Eq.(\ref{EB22}). Furthermore Fig. 7 reveals that switching from the Markovian regime to non-Markovian regime gives rise to the quantum speedup process. It is also seen that both driving field and strong coupling can enhance the non-Markovianty and speed up the evolution of qubit.

\section{Conclusions}
In summary, we investigated quantum evolution of an open moving-qubit modulated by a classical driving field. Firstly, we constructed a model of an open moving-qubit driven by the external classical field, where the environment at zero temperature has the Lorentzian spectral density. Secondly, we obtained an analytical solution of the density operator of this qubit in the dressed-state basis. Thirdly, we analyzed the quantum evolution dynamics by using the QSLT and the non-Markovianity. The results showed that both the non-Markovian environment and the classical driving can speed up the quantum evolution and increase the non-Markovianity in the evolution process, while the qubit motion will delay the quantum evolution and decrease the non-Markovianity. Moreover, the quantum speedup process is induced by the non-Markovianity and the critical points only depend on the qubit velocity. In a way, we can utilize the classical driving to suppress the negative effect of the qubit velocity on the speedup evolution if the qubit is not completely stationary in the experiment. That is to say, the controllable operation of quantum evolution can be realized by adjusting the classical driving strength, the qubit-cavity coupling and the qubit velocity. Also, we give the corresponding physical explanation by using the decoherence rates. The potential candidates that can effectively utilize this approach include cavity QED\cite{Varcoe,McKay}, trapped ions\cite{Jonathan}, superconducting qubits \cite{You.2011,Xiu Gu}.

\section*{Acknowledgments}
This work is supported by the National Natural Science Foundation of China (Grant No.11374096).

\section*{Data Availability Statement}
The datasets generated during and/or analysed during the current study are available from the corresponding author on reasonable request.

\end{document}